\documentclass[year=26,pdfa]{fmcad}

\title{DSLean: A Framework for Type-Correct Interoperability Between Lean 4 and External DSLs}

\author{Tate Rowney, Riyaz Ahuja, Jeremy Avigad, Sean Welleck \\
Carnegie Mellon University, Pittsburgh, PA \\
\{trowney, riyaza, avigad, swelleck\}@andrew.cmu.edu}

\usepackage[T1]{fontenc}
\usepackage[utf8]{inputenc}
\usepackage{amssymb}
\usepackage{amsmath}
\usepackage{algorithm}
\usepackage{algorithmicx}
\usepackage{algpseudocode}

\algrenewcommand\algorithmicindent{0.75em}

\usepackage{listings}
\usepackage{listingsutf8}
\usepackage{xcolor}         

\definecolor{keywordcolor}{rgb}{0.1, 0.2, 0.7}   
\definecolor{tacticcolor}{rgb}{0.0, 0.1, 0.6}    
\definecolor{commentcolor}{rgb}{0.1, 0.5, 0.1}      
\definecolor{symbolcolor}{rgb}{0.0, 0.1, 0.6}    
\definecolor{sortcolor}{rgb}{0.1, 0.5, 0.1}      
\definecolor{stringcolor}{rgb}{0.7, 0.1, 0.1} 
\definecolor{attributecolor}{rgb}{0.7, 0.1, 0.1} 
\definecolor{errorcolor}{rgb}{1.0, 0.0, 0.0}

\pdfoutput=1

\begin{document}

\maketitle

\begin{abstract}

Domain-specific languages (DSLs) mediate interactions between interactive proof assistants and external automation, but translating between the prover's internal representation and such DSLs is a tedious engineering chore. To simplify this task, we present \emph{DSLean}, a framework for bidirectional translation between expressions in the Lean proof assistant and external syntax. DSLean requires only a specification of an external language and its Lean equivalents, abstracting away meta-level implementation details. We demonstrate DSLean's capabilities by implementing three new automation tactics, providing access to external solvers for interval arithmetic, ordinary differential equations, and ring ideal membership.

\end{abstract}
\section{Introduction}

Interactive theorem provers (ITPs) are increasingly used to formally verify claims in mathematics and software. 
The process of verification is often assisted by the use of external automation, either by reconstructing automatically discovered proofs within the proof assistant or using them as a trusted oracle. Communication between the ITP and the external solver is generally mediated by a domain-specific language (DSL) interpretable by both. Examples within Lean 4 \cite{lean4}, a widely used ITP, include interfacing with external SMT solvers and theorem provers \cite{leanauto, leanSMT, canonical}, computer algebra systems \cite{cvxlean, mathematicalean}, and specialized tools for program verification \cite{aeneas, Cproving} and cryptography \cite{CoqCryptography}. 

Implementing such translations is generally challenging. There is little general infrastructure for exporting such information from Lean, so translation often requires the use of bespoke intermediate structures. In the other direction, transforming raw serialized data from an external tool into Lean expressions requires meta-level code to parse the data into a concrete syntax tree, elaborate it into an abstract syntax tree, and modify or recombine it to form a Lean expression. This results in application-specific code that is difficult to maintain and understand, and often requires expertise with Lean metaprogramming. 

\bigskip

\noindent \emph{Contributions.}
We present DSLean, a framework for bidirectional translation between Lean 4 and arbitrary domain-specific languages. DSLean is implemented fully in Lean using its capabilities as a functional programming language, and uses Lean's native parser and metavariable unification algorithm to allow for robust transformations of type-correct Lean expressions to well-formed syntax and vice versa. Despite its internal complexity, DSLean provides a user interface that abstracts away implementation details such as parsing and elaboration.

We also showcase several applications of DSLean, in the form of Lean automation tactics:
\begin{itemize}
    \item The \textbf{\texttt{gappa}} tactic uses DSLean to interface with the Gappa \cite{gappa} interval arithmetic solver, automatically proving nontrivial bounds on real numbers by translating Gappa's proof certificates, emitted in Rocq \cite{rocq} syntax, into Lean.

    \item The \textbf{\texttt{desolve}} tactic connects to the SageMath \cite{sagemath} computer algebra system to obtain general solutions to ordinary differential equations.

    \item The \textbf{\texttt{lean\_m2}} tactic communicates with the Macaulay2 package \cite{macaulay2}, providing witnesses to problems involving ring ideal membership and Gr\"obner bases. 
\end{itemize}
Due to DSLean's capabilities, each example above has a short and interpretable implementation, using comparatively few lines of code (203 for {\texttt{gappa}}, 95 for {\texttt{desolve}}, and 315 for {\texttt{lean\_m2}}). We believe DSLean's general nature makes it applicable to many forms of proof automation beyond what is presented here. 

\section{Overview}

We provide a brief overview of the capabilities of DSLean by showcasing a translation between Lean objects and a small fragment of the Python programming language. DSLean is used by specifying equivalences between Lean objects and external syntax. These are declared using the \texttt{external} keyword, with external syntax on the lefthand side of a \texttt{<==>} delimiter and Lean objects on the right, as shown below:
    \begin{lstlisting}
external translate_Python where
  "True" <==> True
  "False" <==> False
\end{lstlisting}

Translation into Lean may then be performed using the \lstinline{fromExternal} function, while the \lstinline{toExternal} function translates Lean into external syntax (with Lean's \texttt{\#check} command used to show the value and type of these expressions):

\begin{lstlisting}
#check fromExternal translate_Python "True" 
-- Output(*\color{commentcolor}{:}*) (True (*\color{commentcolor}{:}*) Prop)

#check toExternal translate_Python False 
-- Output(*\color{commentcolor}{:}*) ("False" (*\color{commentcolor}{:}*) String)
\end{lstlisting}

More complicated syntax can be expressed by leaving free variables in both sides: 
these are iteratively filled in during translation using the rest of the provided equivalences. For example, the pattern \lstinline{"not" x} shown below will translate syntax beginning with \lstinline{"not"}, then continue to translate the remaining syntax corresponding to \lstinline{x} (and vice versa):

\begin{lstlisting}
external translate_Python where
  ...
  "not" x <==> ¬ x

#check fromExternal translate_Python 
    "not True" 
-- Output(*\color{commentcolor}{:}*) ((*\color{commentcolor}{$\lnot$}*)True (*\color{commentcolor}{:}*) Prop)

#check toExternal translate_Python  
    ¬ False 
-- Output(*\color{commentcolor}{:}*) ("not False" (*\color{commentcolor}{:}*) String)
\end{lstlisting}

As DSLean only translates to and from type-correct Lean expressions, the types of the Lean objects in question impose typing constraints on the external syntax, allowing for representation of complex language specifications. In the previous example, \lstinline{"not" x} will only be translated if the syntax corresponding to \lstinline{x} translates to an object of type \texttt{Prop}. 

DSLean allows specified languages to enjoy levels of polymorphism similar to Lean itself. The addition operator is polymorphic in Lean, and thus so is its translated counterpart:

    \begin{lstlisting}
external translate_Python where
  ...
  a "(*\color{stringcolor}{+}*)" b <==> a + b

  "int(1)" <==> (1 : ℤ)
  "float(1)" <==> (1 : Float)

#check fromExternal translate_Python "int(1) (*\color{stringcolor}{+}*) int(1)" 
-- Output(*\color{commentcolor}{:}*) (1 (*\color{commentcolor}{+}*) 1 (*\color{commentcolor}{:}*) ℤ)

#check fromExternal translate_Python "float(1) (*\color{stringcolor}{+}*) float(1)" 
-- Output(*\color{commentcolor}{:}*) (1 (*\color{commentcolor}{+}*) 1 (*\color{commentcolor}{:}*) Float)

\end{lstlisting}

DSLean also shares meta-level information across translation such as identifier names, whose correspondence may also be specified (as with \lstinline{variableName} below).
Translated free variables are transformed into bound variables when inside relevant binders. 

\begin{center}
    \begin{lstlisting}
external translate_Python where
  ...
  ($variableName) "(*\color{stringcolor}{=}*)" val ";" rest <==> let variableName := val; rest

#check fromExternal translate_Python "myvar (*\color{stringcolor}{=}*) int(1); myvar" 
-- Output(*\color{commentcolor}{:}*) (let myvar (*\color{commentcolor}{:=}*) 1; myvar (*\color{commentcolor}{:}*) Int)
\end{lstlisting}
\end{center}

As some applications may require translations that are not fully reversible, DSLean allows non-injective equivalences to be specified in only one direction. Should an equivalence not be fully reversible (as discussed in Section \ref{sec:LeantoExternal}), DSLean will detect this and inform the user. For example, the following translation is not reversible due to the absence of the variable \lstinline{b} on the righthand side, but functions correctly in the forwards direction:

\begin{center}
    \begin{lstlisting}
external Python_one_way where
  "(" a "," b ")[0]" ==> a

#check fromExternal Python_one_way
    "(True, False)[0]" 
-- Output(*\color{commentcolor}{:}*) (True (*\color{commentcolor}{:}*) Prop)
\end{lstlisting}
\end{center}

Finally, DSLean allows for significant customizability, allowing for manual changes to the associativity or precedence of syntax while automatically determining reasonable values when left unspecified. Further documentation on specific configuration options may be found accompanying the source code. 

\section{Related Work}

Translation between automated theorem provers and domain-specific languages for the purpose of automation is a highly established paradigm. Programs that use such communication to obtain proof certificates are commonplace when interfacing with SMT solvers \cite{leanSMT, smtCoq} or when creating ``hammers'' that rely on first-order theorem provers \cite{leanhammer, leanauto, isabellehammer, coqhammer, agdaprover, duper}. Such programs transform goals in dependent type theory into the propositional or first-order logic syntax usable by the automation they rely on \cite{highertofirstorder}. Additionally, automation that provides witnesses to a statement \cite{Mathlib} or relies on the correctness of established computer algebra systems \cite{mathematicalean, cvxlean} has received growing interest due to ITPs' ability to verify auxiliary reasoning steps performed on such problems. Finally, many program verification frameworks seek to translate syntax of an unverified programming language into a formal specification to allow proofs about its properties \cite{aeneas, cslib, cakeML, compcert, Cproving}. However, translations performed by the above works are highly domain-specific: particularly within the Lean implementations, this results in large amounts of opaque and non-generalizable meta-level code. 

Lean 4 \cite{lean4} possesses a native foreign function interface, as well as an extensible parser and elaborator to define and implement new syntax \cite{leaninterface}, similar to that of Rocq \cite{rocq, rocqFFI}. However, these provide information at the implementation level, often necessitating manual manipulation of intermediate data structures to allow its use in proofs and requiring proficiency in Lean's more obscure implementation details. Work in progress on SciLean \cite{scilean} facilitates communication between certain forms of Lean objects and external representations; however, it requires manual specification of intermediate structures, and does not implement parsing. Additionally, prior work has implemented translation between Rocq and specific protocols used by computer algebra systems \cite{coqCAS}. However, to our knowledge DSLean represents the most comprehensive attempt to simplify and streamline the process of translation between an ITP and arbitrary external DSLs. 

Additionally, we believe the example tactics presented here, while intended as case studies, nonetheless fill niches not occupied by current Lean automation. Interval arithmetic automation in Rocq is highly capable \cite{coqinterval}, explaining our decision to integrate a solver intended for Rocq \cite{gappa}. However, Lean lacks an equally capable equivalent: aside from the in-progress \texttt{lean-cert} project \cite{leancert} which appears to focus primarily on bounding and root-finding for functions, complicated real-valued interval problems are not easily solvable by existing automation. Work on formalization of differential equations in Lean is also sparse, with only minor results in Mathlib \cite{Mathlib} and partial formalization of some numerical methods in SciLean \cite{scilean}; considering this, we believe an oracle-based tactic as showcased here is appropriate, as similar automation has been used in Isabelle/HOL \cite{HOL_ODEs}. Finally, we believe the presented automation for proving ring ideal membership represents an improvement in capability over Lean's \texttt{grind} and Mathlib's \texttt{polyrith}, which do not include capabilities surrounding polynomials over non-standard rings and fields; and an improvement in interpretability over the automation given in Shen et al. \cite{groebner_proj}, which utilizes thousands of lines of bespoke meta-level code. 

\section{DSLean}

We now provide an overview of DSLean's design and implementation. DSLean is implemented in Lean and uses Lean's metaprogramming APIs to perform translation from Lean expressions to external syntax (Section \ref{sec:LeantoExternal}), and from external syntax back to Lean (Section \ref{sec:ExternaltoLean}). Translation into external syntax proceeds by partially matching a target expression with an expression corresponding to a pattern, recursively repeating this process until the entire target is matched. Translation into Lean begins by parsing a raw input into an intermediate data structure (Section \ref{sec:ExternalToLeanParsing}). This data structure is then \textit{elaborated}, or transformed from a representation of the syntax into a data structure representing the expression's semantics, by recursively combining the expressions corresponding to each detected pattern (Section \ref{sec:ExternaltoLeanElaboration}). These processes are described in detail below. 

\subsection{Design Principles}
\label{sec:DesignPrinciples}

DSLean is designed to achieve three goals. First, rather than require the user to specify type- and structure-related constraints in the DSL, we instead utilize Lean's semantics to impose such structure. The types of Lean objects induce typing constraints on corresponding external syntax, allowing for more nuanced structural properties than a simple inductively-defined specification. Lean allows users to define and use objects with a wide variety of useful signatures should available objects be insufficient to induce the desired structure. 

Second, we strive for round-trip consistency; that is, translating an expression out of and back into Lean, or vice versa, should result in the same object modulo semantically irrelevant differences in syntax. As DSLean utilizes Lean's parser and other unsafe or unverifiable code, we provide no formal guarantee of this property. However, apart from initial parsing or final serialization, all processing is performed by substituting equivalent expressions, allowing a higher level of assurance that this property holds. 

Finally, we attempt to make DSLean as simple and intuitive as possible for the user, requiring only the minimum amount of non-optional configuration and allowing them to disregard metaprogramming-related trivialities. Methods of inferring reasonable values for unspecified parameters are described below. 

The above goals still allow for substantial flexibility in the types of language features that DSLean is capable of encoding, including binders, loops, representations of state via monads, and representations of weakly-typed languages using Lean's coercion-related typeclasses. However, it still contains some limitations, including languages whose syntax changes depending on the semantics of previous expressions, as well as those which use indentation to indicate control flow; future work could improve DSLean by allowing these features to be encoded too. 

\subsection{Translation: Lean to External}
\label{sec:LeantoExternal}
Translation from Lean objects to external syntax does not use Lean's default pretty-printing system: rather, we opt to recursively check whether a Lean expression is semantically identical to any equivalences to ensure one-to-one correspondence with external patterns. 

Each user-specified \textit{equivalence}, written via the syntax \texttt{...<==>...}, contains an external pattern on the left in the form of a sequence of strings and identifiers, and a Lean pattern on the right via well-formed Lean syntax. A Lean pattern is stored using Lean's internal representation of expressions, the \texttt{Expr} type. An external pattern consists of a sequence of \textit{parts}, which fall into one of two categories; a \textit{terminal} part can correspond to a literal or the name of a bound variable, while a \textit{nonterminal} part represents a placeholder to be filled by other equivalences. The following example demonstrates how these parts are declared:

\begin{lstlisting}
external simpleExample where

 "terminal_string" <==> 0

 "other_terminal_string" $terminal_binder nonterminal <==> (fun terminal_binder => nonterminal)
\end{lstlisting}
\label{listing:impl_example_1}

The above DSL will translate the Lean value of (for example) \lstinline{fun my_binder_name => 0} by matching with the second equivalence, filling in the \lstinline{$terminal_binder} with the name of the corresponding binder (\lstinline{my_binder_name}), and continuing to match whatever corresponds with \lstinline{nonterminal} (in this case, the function body \lstinline{0}), obtaining the external representation \lstinline{"other_terminal_string my_binder_name terminal_string"}. DSLean \textit{elaborates} the Lean patterns of each equivalence, turning them from raw syntax into expressions, at the time that the equivalences are declared to ensure that terminal binder and nonterminal parts in the external pattern indeed correspond to names of binders and free variables in the Lean pattern; in the above example, this would involve ensuring that the binder \lstinline{terminal_binder} and the free variable \lstinline{nonterminal} indeed exist on both sides. The two sides may have differing numbers of nonterminals if the equivalence is specified as only translating in a single direction (declared with \lstinline{...==>...} or \lstinline{...<==...}). 

\begin{algorithm}
\caption{Translation of a Lean expression to external syntax}

\begin{algorithmic}[1]
\Procedure{LeanToExternal}{$\mathit{leanExpr}$}
    \State $\mathit{out} \gets \text{``''}$ \Comment{Empty string}
    \ForAll{$(\mathit{externalPattern}, \mathit{leanPattern}) \in \mathit{equivalences}$}
        \Comment{sorted so trivial patterns come last}
        \If{\textsc{DefinitionallyEqual(}{$\mathit{leanExpr}, \mathit{leanPattern}$}\textsc{)}}
            \ForAll{$\mathit{part} \in \mathit{externalPattern}$}
                \If{$\mathit{part}$ is a terminal string}
                    \State $\mathit{out} \gets \mathit{out} \mathbin{+\!\!+} \mathit{part}$
                \ElsIf{$\mathit{part}$ is a terminal binder name}
                    \State $\mathit{out} \gets \mathit{out} \mathbin{+\!\!+} \textsc{BinderName}({\mathit{part}, \mathit{leanExpr}} \textsc{)}$
                \ElsIf{$\mathit{part}$ is a nonterminal}
                    \State $\mathit{target} \gets  \textsc{CorrespondingSubexpression(}$
                    ${\mathit{part}, \mathit{leanExpr}}\textsc{)}$
                    \State $\mathit{out} \gets \mathit{out} \mathbin{+\!\!+} \textsc{LeanToExternal(}{\mathit{target}}\textsc{)}$
                \EndIf
            \EndFor
            \State \Return $\mathit{out}$
        \EndIf
    \EndFor
    \State \Call{ThrowError}{``No patterns matched; expression is not translatable''}
\EndProcedure
\end{algorithmic}
\label{alg:LeanToExternal}
\end{algorithm}

The high-level algorithm for using a set of equivalences to translate a Lean expression into external syntax is outlined in Algorithm \ref{alg:LeanToExternal}: it attempts to match a target expression with each equivalence's Lean pattern using Lean's \textit{definitional equality} checker, which reliably determines how to unify semantically equivalent expressions despite syntactic differences. To allow this, the variables in a Lean pattern corresponding to nonterminals are replaced with \textit{metavariables}, syntactic placeholders for as-yet undetermined sub-expressions; asserting definitional equality with a target expression will assign to each metavariable a value among sub-expressions of the target such that the entire pattern matches. This allows easy determination of what sub-expressions in a pattern correspond to nonterminals and require further processing. Upon finding a definitionally equal Lean pattern, the algorithm loops through each part of the relevant external pattern, filling in the external representation and using recursive calls to complete translation of any nonterminals. 

Certain Lean patterns may be general enough such that they match with nearly any input expression without making substantial progress. For example, a Lean pattern such as \lstinline{let myvar := "myval"; nonterminal} will be definitionally equal to any expression by simply ignoring the declaration of \lstinline{myvar}; however, the recursive call on line 12 of Algorithm \ref{alg:LeanToExternal} will pass the same argument as the parent call received modulo reductions (as \lstinline{let myvar := "myval"; nonterminal} is semantically equivalent to \lstinline{nonterminal} when \lstinline{myvar} is not used), causing the algorithm to not terminate. To avoid this scenario, patterns that may trivially match with anything are checked last, giving priority to patterns that allow progress towards translation. It may still be the case that expressions that are untranslatable by a given DSL will match only with trivial patterns and cause such a failure of termination; as there is likely no practical way to detect this, we simply impose an upper limit on the number of recursive calls to prevent stack overflow and automatically reject such inputs.

While the core algorithm is simple, DSLean's implementation relies on several additional implementation details. First, identifiers like free variable names are translated verbatim without requiring a dedicated equivalence; when a binder is translated, its corresponding bound variables are converted into free variables with the correct name so they can be translated. In the above example, an expression such as \lstinline{fun x => x} would be translated to \lstinline{"terminal_string x x"} in this way, matching the second equivalence and filling in the \lstinline{nonterminal} with the name of the (now) free variable \lstinline{x}. 

Furthermore, additional metavariables that do not correspond to nonterminals may be present in Lean patterns (such as the unspecified type of the \lstinline{terminal_binder} argument to the Lean function in the example in \ref{listing:impl_example_1} above). Since each Lean pattern is elaborated in a separate context, metavariables in that pattern are not necessarily functions of any bound variables that may occur in a target expression, which is required by Lean's unification system for said metavariables' value to be dependent on outside binders. To remedy this, should an outside metavariable $?m$ potentially be unified with an expression $e$ with bound variables $(b_1, ... b_n)$ in scope, we perform the substitution $e[(\lambda x_1. \cdots \lambda x_n. ?m) b_1 \cdots b_n / ?m]$ to ensure such dependence can occur. We assert that this remains definitionally equal to the original expression, as unused bound variables disappear from this sub-expression after $\beta$-reduction; thus, these are equivalent when used by Lean's definitional equality checker. 

\subsection{Translation: External to Lean}
\label{sec:ExternaltoLean}

Translation from external syntax to Lean proceeds in two steps: parsing (\ref{sec:ExternalToLeanParsing}) and elaboration (\ref{sec:ExternaltoLeanElaboration}); together, these create a shallow embedding of the specified syntax into Lean. 

\subsubsection{Parsing} 
\label{sec:ExternalToLeanParsing}

External syntax is parsed using Lean's native parser \cite{lean4}, a Pratt parser \cite{prattParser} with additional dynamic information in the style of Swierstra and Duponcheel's combinator parsers \cite{combinatorParsers}. We chose to use it due to its efficient implementation, its robustness as already demonstrated by the above literature, and the fact that it was designed with extensibility in mind, making it suitable for processing a wide variety of reasonably-designed external grammars. 

The external syntax portion of each user-specified equivalence is transformed into a single \textit{production rule}; string literals and free variables in the specification correspond to terminal and nonterminal symbols respectively. 
For binary operations and production rules with more than one nonterminal in a row (for example, a pattern such as \lstinline{"gcd" a b}), the first symbol is given higher precedence to avoid potential interference with highly generic rules such as function application: we do not wish for \lstinline{"gcd" a b} to be interpreted as ``\texttt{gcd} of the function \lstinline{a} applied to \lstinline{b}". This causes operations to be left-associative by default: we believe this is reasonable, as many of Lean's standard mathematical operations are left-associative. Users may modify the associativity of a production rule with the \texttt{+rightAssociative} flag, or manually modify the precedence with the \texttt{precedence := ...} option. 

Lean's parser implements optimized parsing of numerals via a separate parser category: we retain this behavior for performance and convenience, as we believe requiring the inductive specification of the definition of base-10 numerals in every DSL imposes an unnecessary burden on the user. We also use Lean's identifier parser at maximum precedence to parse free variables. 

While powerful, Lean's parser makes certain assumptions about the structure of grammars it parses, causing it to refuse certain production rules. These include rules containing certain reserved terminal symbols such as lone numerals or quotation marks, as well as identifiers that begin with certain characters such as punctuation. It is likely that some or all of these limitations are caused by certain APIs used to expose them and not due to limitations of its theoretical foundations. Future work could address this by modifying or augmenting the parser for greater generalizability. 

\subsubsection{Elaboration}
\label{sec:ExternaltoLeanElaboration}

DSLean does not utilize Lean's default term elaborator directly, instead creating a custom elaboration stage built on top of it. Following the structure given by the parsed syntax tree, the relevant equivalences' Lean patterns are merged into a single expression using Lean's definitional equality checker. 


\begin{algorithm}
\caption{Translation of an external syntax tree to a Lean expression}
\begin{algorithmic}[1]
\Procedure{ExternalToLean}{$\mathit{syntaxTree}$}
    \State $(\mathit{leanPattern}, \mathit{nonterminalTrees}, \mathit{binderNames}) \gets \Call{CorrespondingEquivalence}{\mathit{syntaxTree}}$
    \ForAll{$\mathit{placeholderBinder} \in \mathit{leanPattern}$}
        \State $\mathit{leanPattern} \gets \Call{ReplaceBinder}{\mathit{leanPattern}, \newline \mathit{placeholderBinder}, \mathit{binderNames}[\mathit{placeholderBinder}]}$
    \EndFor
    \ForAll{$\mathit{metavariable} \in \mathit{leanPattern}$}
        \If{\Call{IsNonterminal}{$\mathit{metavariable}$}}
            \State $\mathit{nonterminalContents} \gets \textsc{ExternalToLean(}$
            ${\mathit{nonterminalTrees}[\mathit{metavariable}]}\textsc{)}$
            \If{\Call{Unify}{$\mathit{metavariable}, \mathit{nonterminalContents}$} fails with type mismatch}
                \State \Call{ThrowError}{``External syntax is ill-typed; translation is impossible''}
            \EndIf
        \EndIf
    \EndFor
    \State \Return $\mathit{leanPattern}$
\EndProcedure
\end{algorithmic}
\label{alg:externalToLean}
\end{algorithm}

Algorithm \ref{alg:externalToLean} describes this process; given a syntax tree obtained from parsing (see \ref{sec:ExternalToLeanParsing}), it finds the corresponding Lean pattern and replaces each terminal binder and nonterminal with, respectively, the proper binder name and a recursive call to translate the remainder of the syntax. The result of this recursive call is unified with the parent expression using Lean's definitional equality checker, which automatically assigns other metavariables necessary to resolve typing constraints. This process is illustrated in the following DSL:
\begin{lstlisting}
external simpleExample where

  "one" <==> (1 : ℕ)

  x "plus" y <==> x + y
\end{lstlisting}
Here, the second equivalence contains both metavariables corresponding to the nonterminals \lstinline{x} and \lstinline{y}, but also the type of the addition operation (whether it is adding elements of \lstinline{ℕ}, \lstinline{ℤ}, \lstinline{ℝ}, etc.). When translating external syntax such as \lstinline{"one plus one"}, the algorithm will use the second equivalence, perform recursive calls to resolve both \lstinline{x} and \lstinline{y} to \lstinline{1 : ℕ} and unifies them, assigning the type of the \lstinline{+} operation to \lstinline{ℕ → ℕ → ℕ}; after the assignments are made, the returned expression becomes the expected \lstinline{1 + 1 : ℕ}. 

During elaboration, bound variables in expressions are temporarily transformed into free variables to allow other equivalences to reference outside binders from within their bodies, after which all occurrences of this free variable are turned back into a bound variable. This causes the contexts of some metavariables, which are instantiated before this process, to become out of date; to remedy this, we manually modify any metavariable occurring within a binder $x$ to include $x$ in its context. Although this could potentially cause another instance of the metavariable outside this binder to be assigned to a free variable that is not in scope, we assert this transformation is safe, as such a situation is equivalent to attempting to translate an occurrence of a bound variable with no corresponding binder; both situations will be flagged by Lean's kernel at the end as ill-typed expressions. 

Since Lean's parser processes numerals as a separate category in an optimized fashion (see section \ref{sec:ExternalToLeanParsing}), elaboration of numerals is also handled separately without user specification. By default, numerals are elaborated into natural numbers (Lean's \texttt{Nat} type) with coercion applied wherever possible; this behavior may be changed by specifying the \texttt{numberCast} option. 

Finally, the value of certain expressions such as automation tactics cannot be ascertained until the expression's type is known. Such expressions are treated specially: we disallow nonterminals, and wait until translation time to elaborate them so the rest of the translated expression can be used to determine their expected type.

\section{Case Studies}

We present three applications of DSLean in the form of Lean automation tactics. Each of the examples below utilizes an external solver program to search for solutions to a relevant Lean problem; they were implemented primarily using DSLean, with some additional Lean scripting to register them as tactics. Each tactic requires its relevant solver to be installed to find new proofs; however,  proofs do not rely on the solvers once generated, and can be imported and reused by other users once built. 

\subsection{\texttt{gappa}}

The \texttt{gappa} tactic provides verified Lean proofs of interval bounds on real-valued expressions. It uses DSLean to interface with the Gappa interval arithmetic solver \cite{gappa}; Gappa outputs proof certificates in Rocq \cite{rocq} syntax, which DSLean treats as a DSL to translate these certificates into valid Lean terms. Some examples of its usage are shown below:

\begin{lstlisting}
example (x : ℝ) (y : ℝ) :
  x ∈ Set.Icc 0 1 →
  y ∈ Set.Icc (-1) 1 →
  2 * x + y ∈ Set.Icc (-1) 3 := by
    gappa

example (y : ℝ) :
  y ∈ Set.Icc 0 1 →
  y * y * y ∈ Set.Icc 0 1 := by
    gappa

example (y : ℝ) :
  y ∈ Set.Icc 0 1 →
  y * (1-y) ∈ Set.Icc 0 0.5 := by
    gappa

example (a b c : ℝ) :
  c ∈ Set.Icc (-0.3 : ℝ) (-0.1 : ℝ) ∧
  (2 * a ∈ Set.Icc 3 4 → b + c ∈ Set.Icc 1 2) ∧
  a - c ∈ Set.Icc 1.9 2.05 →
  b + 1 ∈ Set.Icc 2 3.5 := by
    gappa

example (x : ℝ) (y : ℝ) :
  x ∈ Set.Icc 0 1 →
  y ∈ Set.Icc 0 1 →
  x * x * y ∈ Set.Icc 0 1 := by
    gappa

    /- Some goals left to solve:
left (*\color{commentcolor}{:}*) 0 (*\color{commentcolor}{$\le$}*) x (*\color{commentcolor}{$\land$}*) x (*\color{commentcolor}{$\le$}*) 1
right (*\color{commentcolor}{:}*) (0 (*\color{commentcolor}{$\le$}*) y (*\color{commentcolor}{$\land$}*) y (*\color{commentcolor}{$\le$}*) 1) (*\color{commentcolor}{$\land$}*) (0 (*\color{commentcolor}{$\le$}*) x(*\color{commentcolor}{$*$}*)x(*\color{commentcolor}{$*$}*)y (*\color{commentcolor}{$\rightarrow$}*) 1 (*\color{commentcolor}{<}*) x(*\color{commentcolor}{$*$}*)x(*\color{commentcolor}{$*$}*)y)
(*\color{commentcolor}{$\vdash$}*) False -/
    have := right.2 (by nlinarith)
    nlinarith
\end{lstlisting}

We use Mathlib's \cite{Mathlib} \texttt{Set.Icc} definition to encode closed intervals between two real numbers; \texttt{gappa} solves a wide variety of problems involving these\footnote{We wish to credit Gappa \cite{gappa} for the fourth problem above; it is a Lean encoding of an example found therein.} that are otherwise difficult or impossible for existing automation. If \texttt{gappa} cannot fully solve a problem, it leaves (often substantially simpler) subgoals for the user to solve, as with the final example above. 

The tactic supports standard arithmetic operations, logical connectives, anonymous functions, and local declarations, which are similar in both Lean and Rocq. More complex Lean definitions such as closed intervals, set membership, and casting natural numbers to reals are equated with the string representations of definitions within Gappa's dedicated Rocq library. These equivalences are easy to specify with DSLean:
\begin{center}
\begin{lstlisting}
external Gappa_output where
  ...
  -- logical connectives
  x "/\\" y  ==> x ∧ y
  x "\\/" y  ==> x ∨ y
  "not" x    ==> ¬ x
  ...
  -- Gappa equivalents of real numbers
  "Reals.Rdefinitions.R" ==> ℝ
  "Gappa.Gappa_definitions.Float2" x y ==> (IntCast.intCast x : Real) * ((2:Real) ^ (y:Int))
   ...
\end{lstlisting}
\end{center}

Translation from Rocq syntax to Lean includes the large amount of Rocq automation used by Gappa: due to their comparatively narrow scope, we approximately recreate each one in Lean using relevant \texttt{simp} lemmas and various arithmetic tactics, as with the example below.\footnote{The example tactic translation presented here is modified slightly from its appearance in the accompanying implementation, where sequences of tactics have been bundled together as macros to improve the code's readability. } Although nearly all of these goals are relatively simple and can be solved by computationally intensive tactics such as Mathlib's \texttt{nlinarith}, we attempt to close each goal with a variety of simpler heuristics for efficiency. Any subgoals left unsolved by this automation are left for the user to fill manually. 

\begin{lstlisting}
   ...
   -- tactic equivalents
   "Gappa.Gappa_pred_bnd.constant1" a b c ==> by 
        dsimp <;> intros <;> norm_num at * <;> (try constructor) <;> simp_all <;> sorry -- If sorry ends up being used, the goal is left for the user to solve instead
   ...
\end{lstlisting}

Combined, these rules allow translation of Gappa's large Rocq proof terms into Lean. 
Apart from a short macro to clean the syntax and remove comments, this translation is executed entirely by DSLean. A truncated example below shows a Rocq proof certificate emitted from Gappa:

\begin{lstlisting}[language={},basicstyle=\ttfamily\color{black},literate={}]


fun (_y : Reals.Rdefinitions.R) =>
    let f1 := Gappa.Gappa_definitions.Float2 (0) (0) in
    let f2 := Gappa.Gappa_definitions.Float2 (1) (0) in
    ...
    let s1 := p1 /\ s2 in
    ...
    let t1 : p3 := Gappa.Gappa_pred_bnd.constant1 _ i3 _ in
    ...
\end{lstlisting}

This certificate undergoes translation via DSLean into the following Lean proof:

\begin{lstlisting}
fun (_y : ℝ) =>
    let f1 := (0 : ℝ) * 2 ^ (0 : ℤ);
    let f2 := (1 : ℝ) * 2 ^ (0 : ℤ);
    ...
    let s1 := p1 ∧ s2;
    ...\end{lstlisting}
\begin{lstlisting}
    let t1 := (by ...) : t3;
    ...
\end{lstlisting}

While Gappa is best known for its ability to work with floating-point arithmetic, we choose to make use of its capabilities in handling exact interval arithmetic in this case study due to the current lack of developed Lean results concerning floating-point numbers. We emphasize that this does not compromise the soundness of the translated Lean proofs; they do not rely on outside axioms, preventing concerns about roundoff errors and numerical imprecision. 
We believe DSLean will also prove useful in creating automation for floating-point libraries once they are in place.

\subsection{\texttt{desolve}}

The \texttt{desolve} tactic provides witnesses to general solutions of ordinary differential equations. DSLean transforms a target problem into a format readable by the SageMath computer algebra system \cite{sagemath}, as well as translating computed solutions back into Lean.

    \begin{lstlisting}
example : isODEsolution
  (fun x => fun y => deriv y x = 1)
  (fun C _ _ x => x + C) := by
  desolve

example : isODEsolution
  (fun x => fun y => x * deriv y x + y x = 0)
  (fun C _ _ x => C/x) := by
  desolve

example : isODEsolution
  (fun x => fun y => deriv y x + y x = 1)
  (fun C _ _ x => (C + Real.exp x) * (Real.exp (-x))) := by
  desolve

example : isODEsolution
  (fun x => fun y => deriv (deriv y) x + 2 * deriv y x + y x = 0)
  (fun _ K1 K2 x => Real.exp (-x) * (K2 * x + K1)) := by
  desolve
    /- Goals remaining: (*\color{commentcolor}{$\vdash$}*) (fun _ K1 K2 x (*\color{commentcolor}{=>}*) (K1 (*\color{commentcolor}{+}*) K2 (*\color{commentcolor}{$*$}*) x)) (*\color{commentcolor}{$*$}*) Real.exp (-x)) (*\color{commentcolor}{=}*) fun _ K1 K2 x (*\color{commentcolor}{=>}*) Real.exp (-x) (*\color{commentcolor}{$*$}*) (K2 (*\color{commentcolor}{$*$}*) x (*\color{commentcolor}{+}*) K1) -/
  funext
  rw [mul_comm]
\end{lstlisting}

This tactic works primarily with the \texttt{isODEsolution} predicate; \lstinline{isODEsolution f g} represents the assertions that all solutions to $y'(x) = f(x, y(x))$ are of the form $g$ modulo up to three constants of integration. For example, the first problem above proves the assertion that all solutions to $y'(x)=1$ are of the form $x \mapsto x+C$ for an arbitrary $C \in \mathbb{R}$. Once a general solution is obtained, it is instantiated in Lean via DSLean as a function of these constants. Should this not be equivalent to the purported solution in the theorem statement, showing their equivalence is left as a goal for the user (as in the last example above). 

\texttt{desolve} supports equations constructed from a wide variety of Mathlib's differentiable real-valued functions, using DSLean to specify their SageMath representations: 

\begin{lstlisting}
external desolve_out where
   ...
   "e^(" x ")" <== Real.exp x
   "sqrt(" x ")" <== Real.sqrt x
   "sin(" x ")" <== Real.sin x
   "diff(" fn "," var ")" <== deriv fn var
   ...
\end{lstlisting}

Functions of a declared variable in SageMath are translated into anonymous functions in Lean; underscore symbols represent unused constants of integration that are ignored by the Lean function. Some examples are given below:

\begin{lstlisting}
>>> x+1          -- SageMath

fun _ _ _ x =>   -- Lean translation
  x + 1


>>> (_K2*x + _K1)*e^(-x) -- SageMath

fun _ K1 K2 x =>       -- Lean translation
  (K2 * x + K1) * Real.exp(-x)


-- SageMath
>>> -1/120*x^5 + 1/6*x^3 - x + sin(x) + C  

-- Lean translation
fun C _ _ x => 
  (-1/120) * x^5 + 1/6 * x^3 - x + Real.sin(x) + C
\end{lstlisting}

SageMath's differential equations library does not provide any form of proof certificate beyond the solution itself; and moreover, Lean does not yet possess many foundational results on the theory of ODEs. As such, unlike all other examples in this paper, the \texttt{desolve} tactic treats SageMath as an oracle, using an axiom of its assumed correctness to complete the proof, and explicitly informing users of such. Despite this, we believe it serves as an example of DSLean's ability to allow communication between Lean and computer-algebra systems; similar tactics remain useful even without additional axioms, such as those which provide proof witnesses (for example, Mathlib's \texttt{polyrith}).

\subsection{\texttt{lean\_m2}}

The \texttt{lean\_m2} tactic automatically proves statements about membership in ring ideals, a core part of algorithms that use Gr\"obner bases to attack problems in program verification \cite{GrobnerBasisLoopInvariants} and kinematics \cite{GrobnerBasesKinematics}. By communicating with the Macaulay2 computer algebra package \cite{macaulay2} using DSLean, it finds and reconstructs Lean proofs of membership in a wide variety of user-specified fields and rings. Problems are expressed using Mathlib's API (with \lstinline{Ideal.span} representing the ideal generated by a set of elements); some examples of its usage are given below. 

\begin{lstlisting}[literate={∈}{{$\in$}}2 {ℝ}{{$\mathbb{R}$}}2 {→}{{$\to$}}2 {∧}{{$\land$}}2 {⊢}{{$\vdash$}}2 {ℚ}{{$\mathbb{Q}$}}2 {ℤ}{{$\mathbb{Z}$}}2 {ℂ}{{$\mathbb{C}$}}2 {⧸}{{$\backslash$}}2]
example (x y : ℤ) : 2 * x + 3 * y ∈ Ideal.span {x, y} := by lean_m2

example (x y : ℚ) : x^2 * y + y^3 ∈ Ideal.span {x, y} := by lean_m2

example (a b c d e f : ℚ) : a^4+a^2*b*c -a^2*d*e 
    +a*b^3 + b^2*c*d -b^2*e*f + a*c^3 + b*c^2*d -c^2*f^2 
    ∈ Ideal.span {a^2+b*c-d*e, a*b+c*d-e*f, a*c+b*d-f^2} := by lean_m2

\end{lstlisting}

It also supports problems over finite fields, polynomial rings, and quotients of any of these:

\begin{lstlisting}[literate={∈}{{$\in$}}2 {ℝ}{{$\mathbb{R}$}}2 {→}{{$\to$}}2 {∧}{{$\land$}}2 {⊢}{{$\vdash$}}2 {ℚ}{{$\mathbb{Q}$}}2 {ℤ}{{$\mathbb{Z}$}}2 {ℂ}{{$\mathbb{C}$}}2 {⧸}{{$\backslash$}}2]
example (x y : ZMod 5) : x^3 + y^3 ∈ Ideal.span {x + y} := by lean_m2

example (x y : ℂ) : x^2 + y^2 ∈ Ideal.span {x - Complex.I * y} := by lean_m2

example (p q : Polynomial ℤ) : p^2 * q + p * q^2 ∈ Ideal.span {p * q} := by lean_m2

open Polynomial in
example (x y : ℚ[X] ⧸ (Ideal.span {(X:ℚ[X])^2})) : x * y ∈ Ideal.span {x^3, y} := by lean_m2

example (x y : ℚ) (h_sum : x+y = 0) : x^3 + y^3 = 0 := by
  suffices h : x^3 + y^3 ∈ Ideal.span {x + y} by
    simp [h_sum] at h
    exact h
  lean_m2
\end{lstlisting}

Additionally, \texttt{lean\_m2} detects and reports when asserted statements about membership do not hold:

\begin{lstlisting}
example (x y z : ℝ) : x^2 * y + z^3 ∈ Ideal.span {x, y} := by 
    lean_m2
/-
lean_m2(*\color{commentcolor}{:}*) Macaulay2 reports nonzero remainder ((*\color{commentcolor}{|}*) x2^3 (*\color{commentcolor}{|}*)) - element is not in the ideal
-/
\end{lstlisting}

This tactic works with Mathlib's definitions concerning ring and field algebra, using DSLean and auxiliary scripting to turn goals into an imperative script for Macaulay2 to compute. In addition to the equivalences below, DSLean facilitates translation of a variety of more complicated rings and expressions; this vocabulary allows \texttt{lean\_m2} to solve a broader range of problems than existing polynomial reasoning tactics such as \texttt{polyrith} \cite{Mathlib}, including polynomials over quotient rings. 

\begin{center}
\begin{lstlisting}
external M2_out where
   ...
  "ZZ" <== ℤ
  "QQ" <== ℚ
  "RR" <== ℝ
  "CC" <== ℂ
  
  "ii" <== Complex.I
  "a0" <== Polynomial.X -- a polynomial indeterminate

  "(" x ")" <== Polynomial.C x -- constant polynomial

   R "/" x <== R / Ideal.span ({x} : Set R)
   ...
\end{lstlisting}
\end{center}

After translation back into Lean by similar methods, \texttt{lean\_m2} then synthesizes proof based on the output, or reports if the goal is unsatisfiable. 

The \texttt{lean\_m2} tactic was originally prototyped without using DSLean. \footnote{The original version of the tactic is hosted at \url{https://github.com/riyazahuja/lean-m2}, and is not included in the supplementary material. } After substituting DSLean in place of a large bespoke translation script, the number of lines of code in its implementation was decreased by a factor of 5.

\section{Conclusion}

We present DSLean, an extensible and user-friendly framework for bridging between Lean 4 and arbitrary external DSLs. We additionally introduce three Lean automation tactics made using DSLean, which take advantage of communication with external solvers to address a range of nontrivial mathematics problems. 

While automation via external solvers is itself sufficient motivation for the use of such a system, there are undoubtedly other applications of communication between ITPs and external languages. Future research could expand on this work by utilizing similar methods to, for example, translate syntax of unverified programming languages into a form amenable to verification. Additionally, although DSLean uses provably correct definitional equality checks to implement almost every stage of translation, it still relies on some unverifiable code, the chief example being Lean's native parser. Future work could focus on modifying the implementation so that properties about translations such as round-trip consistency could be formally proven. 

\section{Acknowledgements}

Work partially supported by NSF Grant DMS-2434614, DARPA ExpMath Grant HR0011262E028, and a gift from Convergent Research.

\bibliography{main}

@inproceedings{lean4,
  title = {The {Lean} 4 Theorem Prover and Programming Language},
  author = {Moura, Leonardo de and Ullrich, Sebastian},
  year = {2021},
  isbn = {978-3-030-79875-8},
  publisher = {Springer-Verlag},
  address = {Berlin, Heidelberg},
  url = {https://doi.org/10.1007/978-3-030-79876-5_37},
  doi = {10.1007/978-3-030-79876-5_37},
  booktitle = {Automated Deduction – CADE 28: 28th International Conference on Automated Deduction, Virtual Event, July 12–15, 2021, Proceedings},
  pages = {625–635},
  numpages = {11}
}

@article{gappa,
author = {Daumas, Marc and Melquiond, Guillaume},
title = {Certification of bounds on expressions involving rounded operators},
year = {2010},
issue_date = {January 2010},
publisher = {Association for Computing Machinery},
address = {New York, NY, USA},
volume = {37},
number = {1},
issn = {0098-3500},
url = {https://doi.org/10.1145/1644001.1644003},
doi = {10.1145/1644001.1644003},
journal = {ACM Trans. Math. Softw.},
month = jan,
articleno = {2},
numpages = {20}
}

@software{sagemath,
author = {{The SageMath Developers}},
doi = {10.5281/zenodo.8042260},
month = feb,
title = {{SageMath}},
url = {https://github.com/sagemath/sage},
version = {10.10.beta7},
year = {2026}
}

@misc{groebner_proj,
      title={Automated Tactics for Polynomial Reasoning in {Lean} 4}, 
      author={Hao Shen and Junyu Guo and Junqi Liu and Lihong Zhi},
      year={2026},
      eprint={2604.13514},
      archivePrefix={arXiv},
      primaryClass={cs.LO},
}

@software{rocq,
  author       = {{The Rocq Development Team}},
  title        = {The {Rocq} Prover},
  month        = sep,
  year         = 2025,
  publisher    = {Zenodo},
  version      = {9.1.0},
  doi          = {10.5281/zenodo.15149628},
}

@misc{macaulay2,
  author = {Grayson, Daniel R. and Stillman, Michael E.},
  title = {Macaulay2, a software system for research in algebraic geometry},
  howpublished = {Available at \url{http://www2.macaulay2.com}}
}

@inproceedings{Mathlib,
author = {{The Mathlib Community}},
title = {The {Lean} mathematical library},
year = {2020},
isbn = {9781450370974},
publisher = {Association for Computing Machinery},
address = {New York, NY, USA},
url = {https://doi.org/10.1145/3372885.3373824},
doi = {10.1145/3372885.3373824},
booktitle = {Proceedings of the 9th ACM SIGPLAN International Conference on Certified Programs and Proofs},
pages = {367–381},
numpages = {15},
location = {New Orleans, LA, USA},
series = {CPP 2020}
}

@inproceedings{prattParser,
author = {Pratt, Vaughan R.},
title = {Top down operator precedence},
year = {1973},
isbn = {9781450373494},
publisher = {Association for Computing Machinery},
address = {New York, NY, USA},
url = {https://doi.org/10.1145/512927.512931},
doi = {10.1145/512927.512931},
booktitle = {Proceedings of the 1st Annual ACM SIGACT-SIGPLAN Symposium on Principles of Programming Languages},
pages = {41–51},
numpages = {11},
location = {Boston, Massachusetts},
series = {POPL '73}
}

@InProceedings{combinatorParsers,
author="Swierstra, S. Doaitse
and Duponcheel, Luc",
title="Deterministic, error-correcting combinator parsers",
booktitle="Advanced Functional Programming",
year="1996",
publisher="Springer Berlin Heidelberg",
address="Berlin, Heidelberg",
pages="184--207",
isbn="978-3-540-70639-7",
doi={https://doi.org/10.1007/3-540-61628-4_7}
}

@InProceedings{leanSMT,
author="Mohamed, Abdalrhman
and Mascarenhas, Tomaz
and Khan, Harun
and Barbosa, Haniel
and Reynolds, Andrew
and Qian, Yicheng
and Tinelli, Cesare
and Barrett, Clark",
title="{lean-smt}: An {SMT} Tactic for Discharging Proof Goals in {Lean}",
booktitle="Computer Aided Verification",
year="2025",
publisher="Springer Nature Switzerland",
address="Cham",
pages="197--212",
isbn="978-3-031-98682-6",
doi={https://doi.org/10.1007/978-3-031-98682-6_11}
}

@InProceedings{leanauto,
author="Qian, Yicheng
and Clune, Joshua
and Barrett, Clark
and Avigad, Jeremy",
title="{Lean-Auto}: An Interface Between {Lean} 4 and Automated Theorem Provers",
booktitle="Computer Aided Verification",
year="2025",
publisher="Springer Nature Switzerland",
address="Cham",
pages="175--196",
doi={https://doi.org/10.1007/978-3-031-98682-6_10}
}

@article{mathematicalean,
  author  = {Robert Y. Lewis and Minchao Wu},
  title   = {A Bi-Directional Extensible Interface Between {Lean} and {Mathematica}},
  journal = {Journal of Automated Reasoning},
  year    = {2022},
  volume  = {66},
  number  = {2},
  pages   = {215--238},
  month   = may,
  doi     = {10.1007/s10817-021-09611-1},
  url     = {https://doi.org/10.1007/s10817-021-09611-1},
  issn    = {1573-0670}
}

@inproceedings{leanhammer,
title={Premise Selection for a {Lean} Hammer},
author={Thomas Zhu and Joshua Clune and Jeremy Avigad and Albert Q. Jiang and Sean Welleck},
booktitle={The Fourteenth International Conference on Learning Representations},
year={2026},
url={https://openreview.net/forum?id=m04JJNeRK6}
}

@inproceedings{isabellehammer,
author = {Paulson, Lawrence and Blanchette, Jasmin},
year = {2015},
month = feb,
pages = {},
title = {Three Years of Experience with {Sledgehammer}, a Practical Link between Automatic and Interactive Theorem Provers},
booktitle = {Proceedings of the 8th International Workshop on the Implementation of Logics},
doi = {10.29007/tnfd}
}

@inproceedings{smtCoq,
  author    = {Micha{\"e}l Armand and Germain Faure and Benjamin Gr{\'e}goire and Chantal Keller and Laurent Th{\'e}ry and Benjamin Werner},
  title     = {A Modular Integration of {SAT/SMT} Solvers to {Coq} through Proof Witnesses},
  booktitle = {Certified Programs and Proofs (CPP 2011)},
  year      = {2011},
  pages     = {135--150},
  address   = {Kenting, Taiwan},
  month     = dec,
  doi       = {10.1007/978-3-642-25379-9_12},
  publisher = {Springer}
}

@article{coqhammer,
  author  = {{\L}ukasz Czajka and Cezary Kaliszyk},
  title   = {Hammer for {Coq}: Automation for Dependent Type Theory},
  journal = {Journal of Automated Reasoning},
  year    = {2018},
  volume  = {61},
  number  = {1},
  pages   = {423--453},
  month   = jun,
  doi     = {10.1007/s10817-018-9458-4},
  url     = {https://doi.org/10.1007/s10817-018-9458-4},
  issn    = {1573-0670}
}

@inproceedings{cvxlean,
  title={Verified reductions for optimization},
  author={Bentkamp, Alexander and Fern{\'a}ndez Mir, Ramon and Avigad, Jeremy},
  booktitle={International Conference on Tools and Algorithms for the Construction and Analysis of Systems},
  pages={74--92},
  year={2023},
  organization={Springer},
  doi={https://doi.org/10.1007/978-3-031-30820-8_8}
}

@InProceedings{coqinterval,
author="Melquiond, Guillaume",
title="Proving Bounds on Real-Valued Functions with Computations",
booktitle="Automated Reasoning",
year="2008",
publisher="Springer Berlin Heidelberg",
address="Berlin, Heidelberg",
pages="2--17",
isbn="978-3-540-71070-7",
doi={https://doi.org/10.1007/978-3-540-71070-7_2}
}

@article{highertofirstorder,
  author  = {Jia Meng and Lawrence C. Paulson},
  title   = {Translating Higher-Order Clauses to First-Order Clauses},
  journal = {Journal of Automated Reasoning},
  year    = {2008},
  volume  = {40},
  number  = {1},
  pages   = {35--60},
  month   = jan,
  doi     = {10.1007/s10817-007-9085-y},
  url     = {https://doi.org/10.1007/s10817-007-9085-y},
  issn    = {1573-0670}
}

@InProceedings{canonical,
  author =	{Norman, Chase and Avigad, Jeremy},
  title =	{{Canonical} for Automated Theorem Proving in {Lean}},
  booktitle =	{16th International Conference on Interactive Theorem Proving (ITP 2025)},
  pages =	{14:1--14:20},
  series =	{Leibniz International Proceedings in Informatics (LIPIcs)},
  ISBN =	{978-3-95977-396-6},
  ISSN =	{1868-8969},
  year =	{2025},
  volume =	{352},
  URN =		{urn:nbn:de:0030-drops-246128},
  doi =		{10.4230/LIPIcs.ITP.2025.14}
}

@InProceedings{agdaprover,
author="Lindblad, Fredrik
and Benke, Marcin",
title="A Tool for Automated Theorem Proving in {Agda}",
booktitle="Types for Proofs and Programs",
year="2006",
publisher="Springer Berlin Heidelberg",
address="Berlin, Heidelberg",
pages="154--169",
isbn="978-3-540-31429-5",
doi={https://doi.org/10.1007/11617990_10}
}

@InProceedings{duper,
  author =	{Clune, Joshua and Qian, Yicheng and Bentkamp, Alexander and Avigad, Jeremy},
  title =	{Duper: A Proof-Producing Superposition Theorem Prover for Dependent Type Theory},
  booktitle =	{15th International Conference on Interactive Theorem Proving (ITP 2024)},
  pages =	{10:1--10:20},
  series =	{Leibniz International Proceedings in Informatics (LIPIcs)},
  ISBN =	{978-3-95977-337-9},
  ISSN =	{1868-8969},
  year =	{2024},
  volume =	{309},
  publisher =	{Schloss Dagstuhl -- Leibniz-Zentrum f{\"u}r Informatik},
  address =	{Dagstuhl, Germany},
  URN =		{urn:nbn:de:0030-drops-207381},
  doi =		{10.4230/LIPIcs.ITP.2024.10},
}

@article{aeneas,
author = {Ho, Son and Protzenko, Jonathan},
title = {Aeneas: Rust verification by functional translation},
year = {2022},
issue_date = {August 2022},
publisher = {Association for Computing Machinery},
address = {New York, NY, USA},
volume = {6},
number = {ICFP},
url = {https://doi.org/10.1145/3547647},
doi = {10.1145/3547647},
journal = {Proc. ACM Program. Lang.},
month = aug,
articleno = {116},
numpages = {31}
}

@misc{cslib,
      title={{CSLib}: The {Lean} Computer Science Library}, 
      author={Clark Barrett and Swarat Chaudhuri and Fabrizio Montesi and Jim Grundy and Pushmeet Kohli and Leonardo de Moura and Alexandre Rademaker and Sorrachai Yingchareonthawornchai},
      year={2026},
      eprint={2602.04846},
      archivePrefix={arXiv},
      primaryClass={cs.LO},
}

@phdthesis{Cproving,
  author       = {Greenaway, David},
  title        = {Automated proof-producing abstraction of C code},
  month        = aug,
  year         = 2014,
  publisher    = {Zenodo},
  doi          = {10.5281/zenodo.50621},
  url          = {https://doi.org/10.5281/zenodo.50621},
  school           = {CSE, UNSW},
}

@inproceedings{cakeML,
author = {Tan, Yong Kiam and Owens, Scott and Kumar, Ramana},
title = {A verified type system for {CakeML}},
year = {2015},
isbn = {9781450342735},
publisher = {Association for Computing Machinery},
address = {New York, NY, USA},
url = {https://doi.org/10.1145/2897336.2897344},
doi = {10.1145/2897336.2897344},
booktitle = {Proceedings of the 27th Symposium on the Implementation and Application of Functional Programming Languages},
articleno = {7},
numpages = {12},
location = {Koblenz, Germany},
series = {IFL '15}
}

@article{compcert,
author = {Leroy, Xavier},
year = {2009},
month = {07},
pages = {},
title = {Formal Verification of a Realistic Compiler},
volume = {52},
journal = {Communications of the ACM},
doi = {10.1145/1538788.1538814}
}

@phdthesis{leaninterface,
    author       = {Ullrich, Sebastian Andreas},
    year         = {2023},
    title        = {An Extensible Theorem Proving Frontend},
    doi          = {10.5445/IR/1000161074},
    publisher    = {{Karlsruher Institut f{\"{u}}r Technologie (KIT)}},
    pagetotal    = {243},
    school       = {Karlsruher Institut für Technologie (KIT)},
    language     = {english}
}

@software{scilean,
    author = {{Tomáš Skřivan}},
    year = {2025},
    title = {Scientific Computing in {L}ean},
    url = {https://github.com/lecopivo/SciLean}
}

@unpublished{rocqFFI,
  TITLE = {{SerAPI: Machine-Friendly, Data-Centric Serialization for COQ}},
  AUTHOR = {Gallego Arias, Emilio Jes{\'u}s},
  URL = {https://minesparis-psl.hal.science/hal-01384408},
  NOTE = {working paper or preprint},
  YEAR = {2016},
  MONTH = Oct,
  HAL_ID = {hal-01384408},
  HAL_VERSION = {v1},
}

@software{leancert,
    title={{LeanCert}: Verified Computation {\&} Interval Arithmetic},
    year={2026},
    author={Alejandro Radisic},
    url={https://github.com/alerad/leancert},
}

@article{GrobnerBasisLoopInvariants,
title = {Generating all polynomial invariants in simple loops},
journal = {Journal of Symbolic Computation},
volume = {42},
number = {4},
pages = {443-476},
year = {2007},
issn = {0747-7171},
doi = {https://doi.org/10.1016/j.jsc.2007.01.002},
author = {E. Rodríguez-Carbonell and D. Kapur},
}

@article{GrobnerBasesKinematics,
  author    = {Stifter, Sabine},
  title     = {Algebraic methods for computing inverse kinematics},
  journal   = {Journal of Intelligent and Robotic Systems},
  year      = {1994},
  volume    = {11},
  number    = {1},
  pages     = {79--89},
  month     = mar,
  doi       = {10.1007/BF01258295},
  url       = {https://doi.org/10.1007/BF01258295},
  issn      = {1573-0409}
}

@InProceedings{CoqCryptography,
author="Tsai, Ming-Hsien
and Fu, Yu-Fu
and Liu, Jiaxiang
and Shi, Xiaomu
and Wang, Bow-Yaw
and Yang, Bo-Yin",
title="CoqCryptoLine: A Verified Model Checker with Certified Results",
booktitle="Computer Aided Verification",
year="2023",
publisher="Springer Nature Switzerland",
address="Cham",
pages="227--240",
isbn="978-3-031-37703-7",
doi={https://doi.org/10.1007/978-3-031-37703-7_11}
}

@InProceedings{coqCAS,
author="Komendantsky, Vladimir
and Konovalov, Alexander
and Linton, Steve",
title="View of Computer Algebra Data from {Coq}",
booktitle="Intelligent Computer Mathematics",
year="2011",
publisher="Springer Berlin Heidelberg",
address="Berlin, Heidelberg",
pages="74--89",
isbn="978-3-642-22673-1",
doi={https://doi.org/10.1007/978-3-642-22673-1_6}
}

@misc{HOL_ODEs,
      title={Certifying Differential Equation Solutions from Computer Algebra Systems in {Isabelle/HOL}}, 
      author={Thomas Hickman and Christian Pardillo Laursen and Simon Foster},
      year={2021},
      eprint={2102.02679},
      archivePrefix={arXiv},
      primaryClass={cs.LO},
}

\end{document}